\RequirePackage[2020-02-02]{latexrelease}
\documentclass[iop]{emulateapj}
\usepackage{graphicx}
\usepackage{apjfonts}
\usepackage{amsmath} 
\usepackage{natbib}
\usepackage{color}
\usepackage[normalem]{ulem}
\newcommand{\prima}{^{\prime}}

\def\be{\begin{equation}}
\def\ee{\end{equation}}
\def\bea{\begin{eqnarray}}
\def\eea{\end{eqnarray}}

\def\Ereal{{\rm Re\{\tilde{\textbf {E}}^{(\text{t})}\}}}
\def\Eima{{\rm Im\{\tilde{\textbf {E}}^{(\text{t})}\}}}
\newcommand{\logamma}{\ln\left(\frac{(1+\gamma_3)^2+\gamma_4^2}{(1-\gamma_3)^2+\gamma_4^2}\right)}
\newcommand{\logammaa}{\ln\left(\frac{(1+\gamma_3)^2+\gamma_5^2}{(1-\gamma_3)^2+\gamma_5^2}\right)}
\defcitealias{ref:PaperI}{Paper~I}
\defcitealias{ref:PaperII}{Paper~II}
\defcitealias{ref:PaperIII}{Paper~III}
\setcitestyle{notesep={; }}

\slugcomment{Submitted to ApJS}
\shorttitle{IV. Analytical formulation of telecentric etalons}
\shortauthors{Bail\'en, Orozco Su\'arez \& Del Toro Iniesta}

\begin{document}
	\title{On Fabry-P\'erot Etalon-based Instruments \\
	IV. Analytical formulation of telecentric etalons}
	
	\author{F.J. Bail\'en, D. Orozco Su\'arez, and J.C. del Toro Iniesta}
	\affil{Instituto de Astrof\'isica de Andaluc\'ia (CSIC), Apdo. de Correos 3004, E-18080 Granada, Spain}
	\email{fbailen@iaa.es, orozco@iaa.es, jti@iaa.es}

\begin{abstract}
	
	Fabry-P\'erot etalons illuminated with collimated beams have been characterized analytically in detail since their invention. Meanwhile, most of the features of etalons located in telecentric planes have been studied only numerically, despite the wide use of this configuration in astrophysical instrumentation over decades. In this work we present analytical expressions for the transmitted electric field and its derivatives that are valid for etalons placed in slow telecentric beams, like the ones commonly employed in solar instruments. We use the derivatives to infer the sensitivity of the electric field to variations in the optical thickness for different reflectivities and apertures of the incident beam and we compare them to the collimated case. This allows us to estimate the wavefront degradation produced by roughness errors on the surfaces of the Fabry-P\'erot and to establish the maximum allowed RMS value of the cavity irregularities across the footprint of the incident beam on the etalon that ensures diffraction-limited performance. We also evaluate the wavefront degradation intrinsic to these mounts, which is produced only by the finite aperture of the beam and that must be added to the one produced by defects. Finally, we discuss the differences in performance of telecentric and collimated etalon-based instruments and we generalize our formulation to anisotropic etalons. 
	
\end{abstract}


\section{Introduction}\label{sec:general}
Fabry-P\'erot interferometers (etalons) are frequently included in solar magnetographs to carry out the wavelength scanning of spectral lines that are sensitive to magnetic fields. Despite the common use of this technology, there is no consensus among the solar community on their optimum configuration within the instrument in terms of both image quality and spectral performance. So far, two setups have been employed: collimated and telecentric. In a collimated mount, the etalon is located on a pupil plane, thus receiving a collimated beam  from each point of the observed object field (at infinity). This mount offers a better spectral resolution than the telecentric one at the expense of shifting the transmission profile across the field of view (FoV). In addition, incident beams always illuminate the same area of the etalon no matter their direction. This means that individual local defects on the etalon are averaged across the full clear aperture, but they have an impact on the transmitted wavefront over the whole FoV. In telecentric setups the etalon is  very close to a focal plane, whereas the entrance pupil is imaged into infinity. Then, the footprint of the incident beam is much smaller than in the collimated case and local defects are directly mapped onto the detector, thus producing point to point variations of both the transmission and the imaging performance. 

 Defects can be caused by deviations of the homogeneity, uniformity and/or geometry of the etalon with respect to the ideal considerations from which the classical model of a Fabry-P\'erot is derived \citep[e.g.,][]{ref:born}. Usually, defects are originated mainly by departures of the reflecting plates from flatness and parallelism. The impact of such deviations on the transmission profile have been studied in numerous works \citep[e.g.,][]{ref:hill,ref:sloggett,ref:hernandez}. An extensive discussion on the influence of such imperfections in the intensity of the transmitted light was carried out by \cite{ref:PaperI}, hereinafter Paper I. Defects produce  variations in the phase of the transmitted electric field as well. These variations can be understood as errors in the transmitted wavefront and cause a degradation of the imaging performance of the instrument. The first work that addressed the influence of defects on the transmitted wavefront is possibly the one made by \cite{ref:vonderluhe}. Their results refer to strictly monochromatic wavefronts in collimated etalons and represent a worst-case scenario.  \cite{ref:scharmer} considered a more realistic approach that included quasi-monochromatic effects that occur because of the limited resolution of the instruments. Both works suggest a large degradation of the image quality in collimated etalons  and recommend the telecentric configuration  to achieve diffraction-limited performance. Their results are restricted, however, to collimated etalons only. A qualitative discussion on the  impact of defects on image quality for telecentric etalons and a computational method to evaluate their influence on the point-spread-function was presented later in \cite{ref:righini}, but an analytical study similar to the ones presented by \cite{ref:vonderluhe} and \cite{ref:scharmer} for telecentric etalons has not been published yet, up to our knowledge. From our point of view, such an approach would give a valuable insight on the imaging performance of this configuration that would allow a proper comparison with collimated setups, especially when taking into account high frequency errors that can affect the transmitted wavefront even if the footprint of the incident beam on etalon is small. 
 
  There are other sources of image degradation apart from physical defects of the etalon. In particular, imperfections on the illumination of the Fabry-P\'erot can reduce the image quality of the instrument, as evaluated in \citetalias{ref:PaperI}. On the other hand, telecentric setups always suffer from a  characteristic wavelength-dependent apodization of the pupil as seen from the etalon. This effect has an impact on the measured maps of the magnetic field and radial plasma velocities and depends greatly on the $f$-number of the incident beam. The influence of pupil apodization on these mounts  was studied for the first time by \cite{ref:beckers} and recently by \cite{ref:PaperIII}, hereinafter Paper III. The choice of the optimum setup (collimated or telecentric) in a given instrument depends, then, not only on the particular map of defects of the etalon, but also on the  optical parameters, tolerances of the instrument and quality of the etalon, as explained by \cite{ref:righini}. 

This work is the fourth in our series of papers. We derive an analytical expression for the electric field transmitted in telecentric etalons and we investigate the sensitivity of the transmission profile and of the transmitted phase of the electric field to variations of geometry and illumination by taking advantage of the analytical derivatives of the electric field.
 Up to our knowledge, this is the first time the electric field equation is solved analytically for a telecentric configuration.
 Such a solution has many practical advantages, apart from wavefront sensitivity analyses, that are not explored here. One of them would be its possible application in the calibration procedure of telecentric instruments, especially for space-borne magnetographs, whose computational capabilities are very limited. 

We start with the derivation of the analytical expression of the transmitted electric field and its derivatives for telecentric setups (Section \ref{sec:analytical}). We continue with an analysis of the impact of defects on the transmitted wavefront (Section \ref{sec:defects}), as well as the one coming from the intrinsic pupil apodization expected in these mounts (Section \ref{sec:intrinsic}). We discuss the advantages and drawbacks of each configuration in terms of imaging performance (Section \ref{sec:discussion}) and, finally, we generalize our formulation to birefringent etalons in Section \ref{sec:bir}. Section 5 summarizes the main results of the paper and draws some conclusions.

\section{Analytical expressions}
\label{sec:analytical}
In a telecentric configuration the etalon is  located at, or very close to, an intermediate image plane of the instrument, whereas the entrance pupil is set to coincide with the object focal plane (see Fig.~6 in \citetalias{ref:PaperI}). This allows for a homogeneous illumination across the etalon provided that the observed object is uniform. The spectral transmission in this setup broadens and differs from the one corresponding to the collimated case (Eq.~[11] in \citetalias{ref:PaperI}) as a result of the finite aperture of the incident beams, but the passband is kept constant  over the FoV (ideally), unlike in collimated mounts. The transmitted intensity cannot be approximated in this configuration like the average transmission corresponding to collimated beams reaching the etalon with different incidence angles. Instead, coherent superposition of electric fields must be carried out to account for the phase mismatches of rays that propagate along different directions.

In \citetalias{ref:PaperI} we showed that, for a monochromatic plane wave that impinges the etalon, the transmitted electric field, ${\bf E}^{\rm (t)}$, is related to the incident electric field, ${\bf E}^{\rm (i)}$, by
\begin{equation}
{\bf E}^{\rm (t)} = \frac{\sqrt{\tau}}{1 - R}  \, \frac{{\rm e}^{{\rm i} \delta/2} - R \, {\rm e}^{{\rm -i} \delta/2}}{1 + F \sin^{2} (\delta/2)} \, {\bf E}^{\rm (i)},
\label{eq:transmitted_field}
\end{equation}
where $\tau$ is the (intensity) transmission factor of the etalon at normal incidence, $\delta$ is the phase difference between two successively transmitted rays and $F$ is a factor that depends exclusively on the reflectivity of the Fabry-P\'erot surfaces. These three factors are related to several parameters of the etalon, like the surface reflectivity, $R$,   absorption, $A$, refraction index, $n\prima$, and thickness, $h$, but also to the  angle of refraction of the beam, $\theta\prima$. The dependencies are given through the following expressions:

\begin{equation}
\tau=\left(1-\frac{A}{1-R}\right)^2,
\label{eq:tau}
\end{equation}
  \begin{equation}
 F=\frac{4R}{(1-R)^2},
 \end{equation} 
and
 
 \begin{equation}
 \delta\simeq\frac{4\pi}{\lambda}n\prima h\cos\theta\prima.
 \label{eq:delta}
 \end{equation}

In an ideal telecentric configuration in which the chief ray is perpendicular to the etalon across the whole FoV, the transmitted electric field of each individual ray depends only on the radial coordinates of the pupil, $r$, whereas the total transmitted electric field (after integration over the pupil), $\tilde{{\bf E}}^{\rm (t)}$, is given by Eq.~(49) of \citetalias{ref:PaperI}. We can normalize the radial coordinate to the pupil radius of the instrument, $R_{\rm pup}$, and rewrite this equation simply as
\begin{equation}
\label{eq:fieldatzero}
\tilde{{\bf E}}^{\rm (t)} = 2 \, \int_{0}^{1} \varrho\, {\bf E}^{\rm (t)} (\varrho) \; {\rm d}\varrho,
\end{equation} 
where $\varrho\equiv rR^{-1}_{\rm pup}$ .
So far we have presented the electric field of the individual rays as a function of $\delta$, which changes with the refraction angle, $\theta\prima$. The latter depends, in turn, on the incident angle, $\theta$. It is convenient to start using Snell's law in order to change the dependence with $\delta$ in Eq.~(\ref{eq:transmitted_field}) to $r$. Since
\begin{equation}
\label{eq:angles}
\cos \theta\prima = \sqrt{1 - \left( \frac{\sin^{2} \theta}{{n\prima}^{2}} \right)}, 
\end{equation}
and $\theta$ is very small for our cases of interest ($f\#\gg 1$), its sine can be approximated by its tangent (see Fig.~7 in \citetalias{ref:PaperI}) to give
\begin{equation}
\label{eq:costhetaprima}
\cos \theta\prima \simeq \sqrt{1-\frac{\varrho^{2}}{4{n\prima}^{2} (f\#)^2}} \simeq 1-\frac{\varrho^{2}}{8{n\prima}^{2} (f\#)^2},
\end{equation}
where $f\#$ is the $f$-number of the incident beam on the etalon. If we now call 
\begin{equation}
\label{eq:a}
a \equiv \frac{2\pi}{\lambda} n\prima h,
\end{equation}
and 
\begin{equation}
\label{eq:b}
b \equiv \frac{1}{8{n\prima}^2 (f\#)^2},
\end{equation}
we can write
\begin{equation}
\label{eq:deltaab}
\frac{\delta}{2} = a(1-b \varrho^{2}).
\end{equation}
Then, Eq.~(\ref{eq:transmitted_field}) can be cast as
\begin{equation}
\label{eq:vectorelectric6}
\begin{gathered}
{\bf E}^{\rm (t)} =
\frac{\sqrt{\tau}}{1-R} \, \bigg[\frac{(1-R)\cos\left(a[1-b \varrho^{2}]\right)}{1+F \sin^{2} \left(a[1-b \varrho^{2}]\right)} \ +\\
{\rm i}\frac{(1+ R)\sin (a[1-b \varrho^{2}])}{1+F \sin^{2} (a[1-b \varrho^{2}])} \bigg] \, {\bf E}^{\rm (i)}.
\end{gathered}
\end{equation}
Now we can write Eq. (\ref{eq:fieldatzero}) as
\begin{equation}
\begin{gathered}
\tilde{{\bf E}}^{\rm (t)} =
2\sqrt{\tau} \left\{
 \int_{0}^{1} \frac{\varrho \cos\left(a[1-b \varrho^{2}]\right)}{1+F \sin^{2} \left(a[1-b \varrho^{2}]\right)} \, {\rm d}\varrho \ + \right. \\
\left. {\rm i} \frac{1+R}{1-R} \, \int_{0}^{1} \frac{\varrho \sin \left(a[1-b \varrho^{2}]\right)}{1+F \sin^{2} \left(a[1-b \varrho^{2}]\right)} \, {\rm d}\varrho \right\} {\bf E}^{\rm (i)}.
\end{gathered}
\label{eq:fieldatzero3}
\end{equation}
This equation has analytical integration. Indeed,

\begin{equation}
\begin{gathered}
\int_{0}^{1} \frac{\varrho \cos\left(a[1-b \varrho^{2}]\right)}{1+F \sin^{2} \left(a[1-b \varrho^{2}]\right)} \, {\rm d}\varrho=\\
\frac{1}{\alpha_1}\left[\arctan\left(\gamma_1\right)-\arctan\left(\gamma_2\right)\right],
\end{gathered}
\label{eq:int_real}
\end{equation}
and
\begin{equation}
\begin{gathered}
\int_{0}^{1} \frac{\varrho \sin \left(a[1-b \varrho^{2}]\right)}{1+F \sin^{2} \left(a[1-b \varrho^{2}]\right)} \, {\rm d}\varrho=\\
\frac{1}{\alpha_2}\left[\ln\left(\frac{(1+\gamma_3)^2+\gamma_4^2}{(1-\gamma_3)^2+\gamma_4^2}\right)-\ln\left(\frac{(1+\gamma_3)^2+\gamma_5^2}{(1-\gamma_3)^2+\gamma_5^2}\right)\right],
\end{gathered}
\label{eq:int_imag}
\end{equation}
where we have defined
\begin{align}
\begin{gathered}
\alpha_1\equiv 2ab\sqrt{F},\\
\alpha_2\equiv 2\alpha_1\sqrt{F+1},\\
\gamma_1\equiv\sqrt{F}\sin a,\\
\gamma_2\equiv\sqrt{F}\sin\left(a[1-b]\right),\\
\gamma_3\equiv\sqrt{\frac{F}{F+1}},\\
\gamma_4\equiv\frac{\tan{\left(\cfrac{a}{2}\left[1-b\right]\right)}}{\sqrt{F+1}},\\
\gamma_5\equiv\frac{\tan{\left(a/2\right)}}{\sqrt{F+1}}.
\end{gathered}
\label{eq:params}
\end{align}

\begin{figure}
	\centering
	\includegraphics[width=\linewidth]{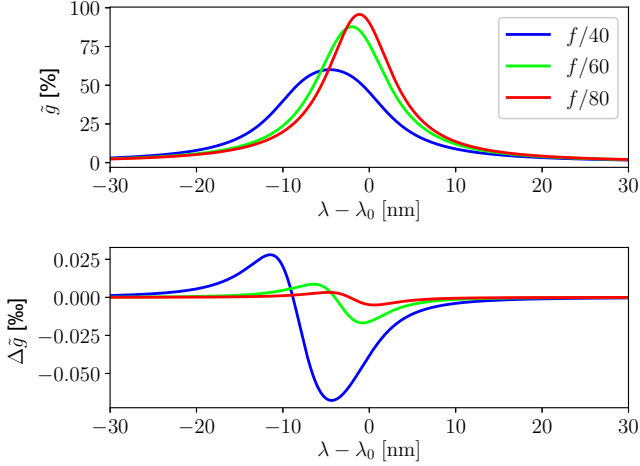}
	\caption{Top: transmission profiles (expressed in $\%$) of an etalon in telecentric beams with $f$-numbers $f/40$ (blue), $f/60$ (green) and $f/80$ (red). Bottom: difference between the transmission profile calculated numerically and the one obtained with the analytical expressions (expressed in \textperthousand).}
	\label{fig:intensitycomp}
\end{figure}

Equations (\ref{eq:int_real})-(\ref{eq:params}) are tedious and, unfortunately, there is no easy way to simplify them further. The reason for this is that they cannot be either expanded into power series or neglected as a result of a large sensitivity of the transmitted electric field to small changes in any of the parameters. Note however that, apart from a transmission factor $\sqrt{\tau}$, the final expression depends only on three coefficients: $R$ (or, equivalently, $F$), $a$ and $b$. This means that the electric field transmitted by the etalon is determined uniquely by the absorptivity and reflectivity of the etalon and by the quantities $n\prima h\lambda^{-1}$ and $n\prima f\#$. Note that the refraction index acts only as an amplification factor of both the thickness and the $f$-number in the equations. Crystalline etalons can benefit, then, from much faster apertures (and, hence, from much smaller etalon and instrument dimensions) while keeping the same spectral and imaging properties, which makes them appropriate in instruments with stringent size requirements and, in particular, in balloon- or space-borne telescopes \citep[e.g.,][]{ref:imax,ref:solanki}. 

Once we have an analytical solution for the transmitted electric field, we can calculate other physical quantities of interest, like the transmission profile of the etalon. Transmission, $\tilde{g}$, is defined as the average ratio between the transmitted and incident intensities in the telecentric configuration and is given simply by

\begin{equation}
\label{eq:transratio2}
\tilde{g} = \frac{\tilde{{\bf E}}^{\rm (t)} {\tilde{\bf E}}^{{\rm (t)} \ast}} {{\bf E}^{\rm (i)} {\bf E}^{{\rm (i)} \ast}}=\frac{{\rm Re}\{\tilde{{\bf E}}^{\rm (t)}\}^2+{\rm Im}\{\tilde{{\bf E}}^{\rm (t)}\}^2} {{\bf E}^{\rm (i)} {\bf E}^{{\rm (i)} \ast}}, 
\end{equation}
where the asterisk indicates the complex conjugate. Figure~\ref{fig:intensitycomp} shows the transmission profile for three incident telecentric beams with $f/40$, $f/60$ and $f/80$ on a crystalline etalon with  $n=2.3$, $h=250$ $\mu$m, $R=0.92$, and $A=0$. We will keep the same parameters for the numerical examples hereinafter. 
Note that the transmission is broadened and shifted to the blue with respect to that of an equivalent collimated configuration tuned at  $\lambda_0=617.3$ nm, as described in detail in \citetalias{ref:PaperI}. The profiles have been calculated with  Equations~(\ref{eq:fieldatzero3}) - (\ref{eq:transratio2}). The differences between these profiles and the ones calculated by numerical integration of Eq.~(\ref{eq:fieldatzero}) are also shown (in \textperthousand). Numerical integration has been carried out without expanding $\cos\theta\prima$ into a power series and both the absolute and relative tolerance for the integration method have been adjusted to be several order of magnitudes stricter than the maximum difference found for each profile. Hence, the tiny differences that appear are basically due to the small angle approximation used to obtain the analytical solution.

The results shown for a crystalline etalon in Fig.~\ref{fig:intensitycomp} and henceforth correspond also to an air-gapped Fabry-P\'erot with the same reflectivity and absorptivity, but a cavity  2.3 times larger ($575$ $\mu$m) and apertures 2.3 times smaller. Table \ref{tab:fnumbers} shows the equivalence between the $f$-numbers employed in the numerical examples presented hereinafter for a crystalline etalon and those corresponding to an air-gapped etalon with $h=575$ $\mu$m. Note that the range of apertures used in our numerical examples is compatible with the $f$-numbers commonly employed in ground-based instruments that use air-gapped Fabry-P\'erots.

\begin{table}
	\begin{center}
		\caption{Apertures employed for the crystalline etalon and the ones corresponding to its equivalent air-gapped etalon.}
		\label{tab:fnumbers}
		\vspace{0.2truecm}
		\begin{tabular}{lcrrr}
			\hline 
			 Crystalline & $f/40$ & $f/60$ & $f/80$ & $f/100$\\
			 Air & $f/92$ & $f/138$ & $f/184$ & $f/230$\\
			\hline
		\end{tabular}
	\end{center}
\end{table} 
 
 Having access to the analytical expression of the transmitted electric field has numerous advantages. For instance, we can calculate the analytical derivatives with respect to $a$ and $b$ to evaluate the sensitivity to variations of any of the etalon parameters. We are going to focus here on the derivative with respect to $a$, because we are interested on the variations on the electric field that arise from changes in thickness across the aperture. The derivative of the transmission profile with respect to $a$ can be cast simply as

\begin{equation}
\begin{gathered}
\frac{\partial \tilde{g}}{\partial a} = 
\frac{2} {{\bf E}^{\rm (i)} {\bf E}^{{\rm (i)} \ast}}
\left({\rm Re}\{\tilde{{\bf E}}^{\rm (t)}\}\frac{\partial}{\partial a}{\rm Re}\{\tilde{{\bf E}}^{\rm (t)}\}+{\rm Im}\{\tilde{{\bf E}}^{\rm (t)}\}\frac{\partial}{\partial a}{\rm Im}\{\tilde{{\bf E}}^{\rm (t)}\}\right).
\end{gathered}
\label{eq:derivative_trans}
\end{equation}
The derivatives of the real and imaginary parts of the electric field are given by

\begin{widetext}
	\begin{equation}
	\begin{gathered}
	\frac{\partial\Ereal}{\partial a}=
	\frac{2\sqrt{\tau}}{\alpha_1}\left[\frac{\alpha_1\prima}{\alpha_1}\left(\arctan\gamma_2-\arctan\gamma_1\right)+\frac{\gamma_1\prima}{1+\gamma_1^2}-\frac{\gamma_2\prima}{1+\gamma_2^2}\right],
	\end{gathered}
	\label{eq:deriv_real}
	\end{equation}
	
	\begin{equation}
	\begin{gathered}
	\frac{\partial\Eima}{\partial a}=
	\frac{2\sqrt{\tau}}{\alpha_2 }\frac{1+R}{1-R}\\
	\times\left[\frac{\alpha_2\prima}{\alpha_2}\bigg\{\logammaa-\logamma\bigg\}+\frac{8\gamma_3\gamma_5\gamma_5\prima}{\left[(1+\gamma_3)^2+\gamma_5^2\right]\left[(1-\gamma_3)^2+\gamma_5^2\right]}-\frac{8\gamma_3\gamma_4\gamma_4\prima}{\left[(1+\gamma_3)^2+\gamma_4^2\right]\left[(1-\gamma_3)^2+\gamma_4^2\right]}\right],
	\end{gathered}
	\label{eq:deriv_ima}
	\end{equation}	
\end{widetext}
where the prime denotes the partial derivative with respect to $a$:
\begin{equation}
\begin{gathered}
\alpha_1\prima=2b\sqrt{F},\\
\alpha_2\prima=4b\sqrt{F(F+1)},\\
\gamma_1\prima=\sqrt{F}\cos a,\\
\gamma_2\prima=\sqrt{F}(1-b)\cos\left(a[1-b]\right),\\
\gamma_4\prima=\frac{1-b}{2\sqrt{F+1}}\sec^2\left(\frac{a}{2}[1-b]\right),\\
\gamma_5\prima=\frac{1}{2\sqrt{F+1}}\sec^2\left(\frac{a}{2}\right).
\end{gathered}
\end{equation}

\begin{figure}
	\centering
	\includegraphics[width=\linewidth]{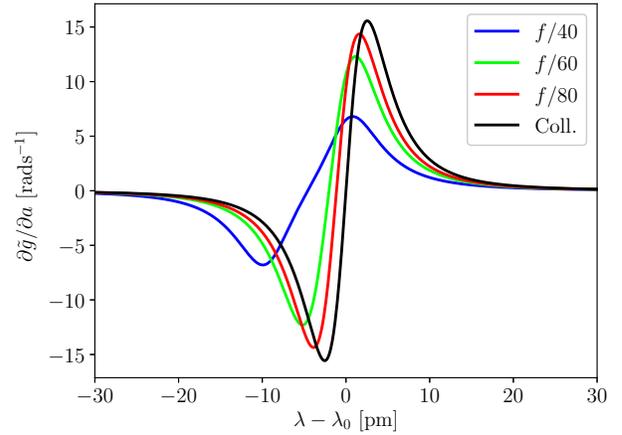}
	\caption{Spectral dependence of the derivative of the transmission profile with respect to $a$ corresponding to a telecentric  $f/60$ mount (blue) and to a collimated configuration (red).}
	\label{fig:deriv_prof}
\end{figure}

Figure \ref{fig:deriv_prof} shows $\partial \tilde{g}/\partial a$ over the transmission profile for both a collimated and three telecentric configurations with different $f$-numbers illuminating the crystalline etalon. Apart from the blue shift that appears also in Fig.~\ref{fig:intensitycomp}, the amplitudes of the telecentric profiles decrease with increasing apertures because the spectral resolution of the etalon worsens with smaller $f$-numbers, which translates into a less steep profile. Each curve is also shifted by the same amount it does its transmission peak. 

 The derivatives of Fig. \ref{fig:deriv_prof}  have an anti-symmetric shape and change of sign at the peak transmission wavelength. Hence, the effect of a change in $a$ is mostly seen as a shift in $\tilde{g}$ and have a negligible impact on its width, as already evaluated in \citetalias{ref:PaperI} for the collimated configuration. Then, impurities and inhomogeneities of the etalon across the footprint of the incident beam produce only different shifts of the transmission profiles point to point. The average effect of the local shifts of the transmission across the footprint of the beam is a broadening of the passband and a reduction of the transmission peak.

\begin{figure}
	\centering
	\includegraphics[width=\linewidth]{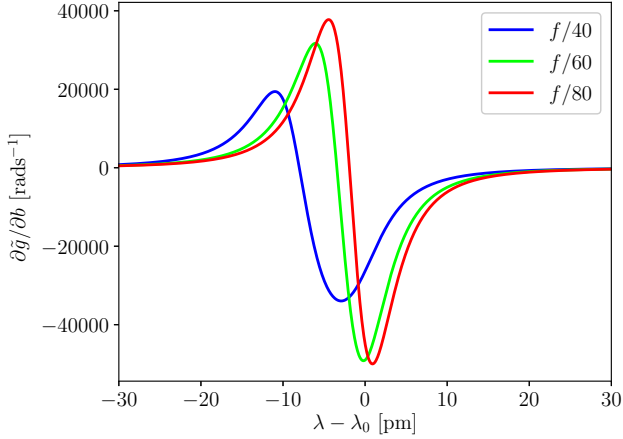}
	\caption{Spectral dependence of the derivative of the transmission profile with respect to $b$ corresponding to telecentric beams with $f/40$ (green), $f/60$ (blue) and $f/80$ (red).}
	\label{fig:deriv_prof_b}
\end{figure}

Note that $\partial \tilde{g}/\partial a$ encodes the sensitivity of the transmission profile to variations of  wavelength and optical thickness, but not to changes of the focal length, which are contained in the partial derivative with respect to $b$ (Eq.~\ref{eq:b}). Fluctuations in the refraction index have an impact on both $a$ and $b$. The derivatives of $\tilde{g}$ with respect to $b$ can easily be calculated by substituting the quantities with primes in Eqs.~(\ref{eq:deriv_real}) and (\ref{eq:deriv_ima}) with their corresponding partial derivatives with respect to $b$. Figure \ref{fig:deriv_prof_b} shows the derivative of the transmission profile with respect to $b$ for different f-numbers. The profiles exhibit a similar behavior, except for a change of sign, to those of Fig. \ref{fig:deriv_prof}, but are not completely anti-symmetric since increasing $b$ has the effect of both shifting the profile and broadening it. The derivatives with respect to $b$ are three orders of magnitude larger, too.  However, $a\sim n\prima h\lambda^{-1}$ is typically about $10^3-10^4$ rads in the etalons employed in solar instruments, whereas $b\sim (n\prima f\#)^{-2}$ is of the order of $10^{-4}-10^{-6}$ rads. In our numerical example, $a$ is approximately nine orders of magnitude larger than $b$, which means that a small perturbation of the former has much more importance on $\delta$ (Eq.~{\ref{eq:deltaab}}) and, hence, on the transmission profile.

\section{Phase error amplification and image quality}
\subsection{Errors introduced by defects}\label{sec:defects}
The transmission profile derived in Sect.~\ref{sec:analytical} corresponds to an ideal homogeneous and isotropic etalon whose reflecting surfaces are perfectly parallel to each other. In a real case, the etalon presents irregularities and/or inhomogeneities that disturb the transmitted electric field and degrade not only the spectral resolution, but also the imaging performance of the instrument.  \cite{ref:righini}   qualitatively discuss the image quality degradation produced in telecentric etalons and provide a method to evaluate the impact when a particular map of the defects is measured. However, a quantitative evaluation of the impact produced by defects on the wavefront in telecentric etalons has not been presented so far, to our knowledge.

Here we will follow \cite{ref:scharmer} to estimate the wavefront error introduced by defects of the etalon in a telecentric mount. This is as simple as calculating the perturbation produced in the optical phase by such irregularities and/or inhomogeneities. We will consider here the sensitivity of the phase to variations of the parameter $a$, $\Delta a$. We neglect for simplicity any variations on $b$ produced by changes of the $f$-number and/or the refraction index, as justified in Section~\ref{sec:analytical}. The distorted phase, $\phi$, can be approximated at first order, then, as
\begin{equation}
\phi\approx \phi_0+\frac{\partial \phi_0}{\partial a}\Delta a,
\label{eq:propagation}
\end{equation}
where $\phi_0$ is the unperturbed ideal phase given by
\begin{equation} 
\phi_0=\arctan\left(\frac{\Eima}{\Ereal}\right),
\end{equation}
which can be evaluated simply using Equations~(\ref{eq:fieldatzero3}),~(\ref{eq:int_real}), and~(\ref{eq:int_imag}). The derivative of the phase can be calculated analytically through the derivatives of the real and imaginary parts of the electric field as

\begin{equation}
\frac{\partial \phi_0}{\partial a}=\cfrac{1}{1+\tan^2\phi_0}\frac{1}{\Ereal}\left[\frac{\partial }{\partial a}\Eima-\tan\phi_0\frac{\partial }{\partial a}\Ereal\right],
\end{equation}
where the derivatives of the electric field are given by Equations~(\ref{eq:deriv_real}) and (\ref{eq:deriv_ima}).
The final expression is cumbersome and will not be presented here, but approaches, for very large $f$-numbers, to 
\begin{equation}
 \lim_{f\# \to \infty}\frac{\partial \phi_0}{\partial a}=\frac{1+R}{1-R} \cfrac{\sec^2 a}{1+\left(\cfrac{1+R}{1-R}\right)^2\tan^2 a},
 \label{eq:lim_deriv}
\end{equation}
which corresponds, as it should be, to the derivative of the phase of the transmitted electric field in an ideal collimated configuration. Note that this equation differs from the one given by \cite{ref:scharmer} in terms of $\delta$ because of the inclusion in our work of a global phase, usually unimportant, on the transmitted electric field (see \citetalias{ref:PaperI} for further details). 

\begin{figure}
 	\centering
 	\includegraphics[width=1.08\linewidth]{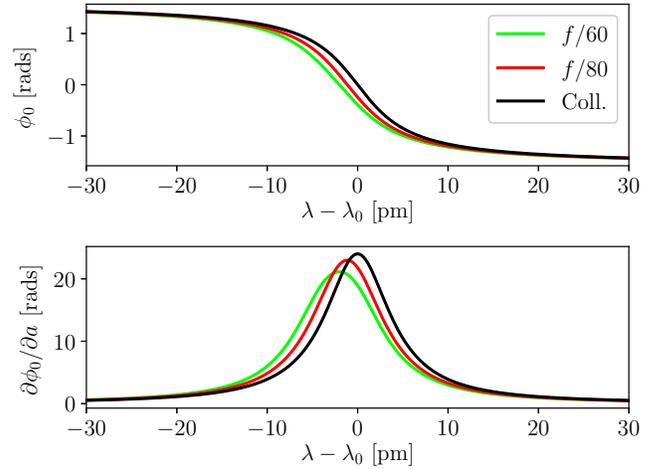}
 	\caption{Top: dependence of the phase of the transmitted field with wavelength across the transmission profile for an $f/60$ telecentric (blue) and a collimated (red) configuration. Bottom: corresponding derivatives of the phase with respect to $a$ across the transmission profile.}
 	\label{fig:phase_wvl}
 \end{figure}

 Figure \ref{fig:phase_wvl} shows the dependence of the phase over the transmission profile for two telecentric beams and for the collimated configuration (top), as well as the corresponding derivatives with respect to $a$ (bottom). The shape and magnitude of the derivatives is quite similar for the collimated and telecentric cases, and so will be the sensitivity of the phase to errors in the optical thickness. Note that the peaks of the derivative in the telecentric cases are shifted to the blue with respect to $\lambda_0$ in the same way it does their transmission profile. Moreover, the derivative reaches higher peaks as the $f$-number is increased. The reason for this is that the the spectral resolution is improved as the $f$-number increases, which translates into a sharper profile of the phase around the transmission peak. 
 
 Again, the results  for the collimated mount do not coincide exactly with the phase error amplification function given by \cite{ref:scharmer} because of the omission in his work of a global phase.\footnote{ The phase error amplification function described by \cite{ref:scharmer} is calculated as the derivative of the phase with respect to $\delta$, instead of to $a$. It is  necessary, then, to include  a factor 2 in his expressions for comparison purposes.} In particular, his expression tends to negative values at wavelengths far from the maximum transmission, whereas ours is always positive. The maximum of the derivative for the collimated configuration is also different: $(1+R)(1-R)^{-1}$ in our case, to be compared with the value of $2R(1-R)^{-1}$ found by \cite{ref:scharmer}. According to our results, the degradation of the wavefront for the collimated mount at the peak transmission is expected to be, then, a bit more optimistic than the one calculated by \cite{ref:scharmer}, especially for low to moderate reflectivities.

\begin{figure}
	\centering
	\includegraphics[width=1.07\linewidth]{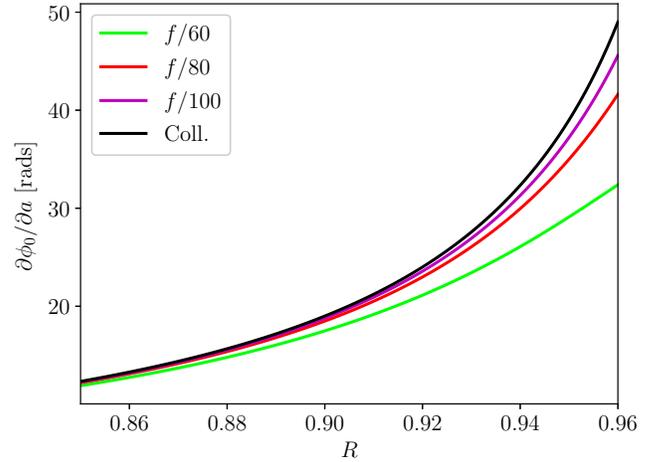}
	\caption{Derivative of the phase with respect to parameter $a$, at the peak transmission wavelength, as a function of the reflectivity of the etalon surfaces for an $f/60$ (green), an $f/80$ (red) and an $f/100$ (magenta) telecentric beam, as well as for the collimated case (black)}
	\label{fig:defects}
\end{figure}

Apart from the aperture of the beam, reflectivity plays also an important role in the degradation of the wavefront error. In particular, the closer the reflectivity to unity the sharper the electric field module and phase profiles. Hence, the derivatives across the transmission profile also increase with larger reflectivities. Figure \ref{fig:defects} shows the derivative of the phase with respect to $a$ as a function of the reflectivity. The derivative has been evaluated at the maximum of the transmission profiles for different apertures of the beam. The collimated case is also shown for comparison purposes. Note that, the larger the $f$-number, the higher the value of the derivative and the more important the impact of the reflectivity on the sensitivity to defects. Once again, the reason for this is that the transmission and phase profiles also get steeper when approaching to collimated illumination. 

\begin{figure}
	\centering
	\includegraphics[width=1.07\linewidth]{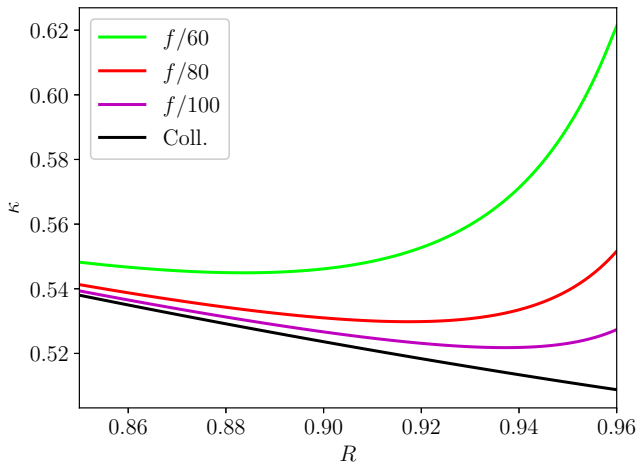}
	\caption{Factor $\kappa$ as a function of the reflectivity for different telecentric beams: $f/60$  (green), $f/80$ (red) and $f/100$ (magenta). The collimated case is also shown (black).}
	\label{fig:phase_effective}
\end{figure}

As expected from Fig.~\ref{fig:phase_wvl}, the wavefront degradation is maximum at the peak of the transmission profile. Hence, the results shown in Fig.~\ref{fig:defects} represent a worst case scenario if used to evaluate the optical quality of the etalon. \cite{ref:scharmer} suggested  that the effect of the finite width of the passband of the etalon can be estimated by multiplying the monochromatic wavefront error produced at the peak transmission by a factor $1/2$. The choice of this factor is not entirely justified, though. A more appropriate approach would consist in calculating the quasi-monochromatic (``effective'') wavefront degradation after integrating the derivative of the phase with respect to $a$ across the transmission profile as

\begin{equation}
\bigg(\frac{\partial \phi_0}{\partial a}\bigg)_{\rm eff}=\cfrac{\displaystyle\int \cfrac{\partial \phi_0}{\partial a} (\lambda)\tilde{g}(\lambda)\,\, {\rm d}\lambda}{\displaystyle\int \tilde{g}(\lambda)\,\, {\rm d}\lambda}.
\label{eq:effective_derivative}
\end{equation}
Of course, we can always relate the quasi-monochromatic derivative of the phase to the monochromatic derivative at the maximum of the transmission profile through a factor $\kappa$
as
\begin{equation}
\bigg(\frac{\partial \phi_0}{\partial a}\bigg)_{\rm eff}=\kappa\bigg(\frac{\partial \phi_0}{\partial a}\bigg)_{\rm peak},
\label{eq:kappa}
\end{equation}
where $\kappa$ depends, in general, on the parameters of the etalon and on its illumination. Figure \ref{fig:phase_effective} shows the dependence of the factor $\kappa$ against the reflectivity for telecentric configurations with different $f$-numbers, as well as for the collimated case.  Our results show that $\kappa$ is very close to $1/2$, as estimated qualitatively by \cite{ref:scharmer}. This is particularly true for collimated mounts and for telecentric setups with ``large'' $f$-numbers ($f\#\geq80$), almost independently of the reflectivity of the etalon. For mounts illuminated with faster beams, $\kappa$ shows a stronger dependence with the reflectivity and amounts to $\sim0.60$  for an $f/60$ beam and $R=0.96$.

Let us imagine now that the thickness of the etalon varies across the footprint of the incident beam following a certain distribution with an RMS value $\langle\Delta h\rangle$. Using Eq.~(\ref{eq:propagation}), the RMS of the effective wavefront error induced by the etalon, $\langle\Delta \phi\rangle_{\rm eff}$, is simply given by

\begin{equation}
\langle\Delta \phi\rangle_{\rm eff}=\bigg|\frac{\partial \phi_0}{\partial a}\bigg|_{\rm eff}\!\!\! \frac{2\pi n\prima}{\lambda}\langle\Delta h\rangle.
\label{eq:phi_rms}
\end{equation}
According to the Mar\'echal criterion, diffraction-limited performance of the etalon is achieved when the RMS of the wavefront degradation keeps below $\lambda/14$.  Using Eq.~(\ref{eq:kappa}), the condition for the RMS of the error thickness to ensure diffraction-limited optical quality, $\langle\Delta h\rangle_{\rm diff}$, can be written, then, simply as
\begin{equation}
\langle\Delta h\rangle_{\rm diff}<\cfrac{\lambda}{14\kappa n\prima\bigg|\cfrac{\partial \phi_0}{\partial a}\bigg|_{\rm peak}}.
\label{eq:rms}
\end{equation}
For a value of the reflectivity of $0.90$, the RMS of the irregularities on the thickness must be better than $\sim\lambda/300$ or $\sim2 $ nm at $\lambda =617$ nm for both a collimated and an $f/60$ configuration. If the reflectivity is increased up to $0.95$, the flatness shall be $\sim\lambda/630$ in the collimated case and $\sim\lambda /550$ in the telecentric configuration to accomplish diffraction-limited performance. The flatness requirement over the footprint of the telecentric setup converges with increasing $f-$number to that of the collimated mount even for large reflectivities. 
  A similar expression to Eq.~(\ref{eq:rms}) can be found for the requirement in homogeneity on the refraction index. If we ignore perturbations introduced by $b$, the maximum allowed RMS of the refraction index variations across the footprint can be shown to be of the order of $\sim 5\cdot10^{-4}$ $\%$ to fulfill the diffraction limit requirement. 
  
  These requirements on the roughness and index refraction homogeneity apply to the area illuminated by the incident beam on the etalon.
  In collimated mounts, where the etalon is at a pupil plane, the whole clear aperture of the etalon is always illuminated no matter the observed point on the object field. In telecentric setups, the footprint of the incident beam is much smaller because of the very close location of the etalon with respect to the image plane. 
On the other hand, the RMS value of thickness errors tend to increase with the aperture, especially if they are caused by one of the large scale defects mentioned in \citetalias{ref:PaperI} (departure of parallelism, spherical defect, sinusoidal defect, etc.). The incident wavefront is usually expected, then, to be much less distorted by defects in telecentric mounts than in collimated setups even though the sensitivity to errors in the thickness is very similar in both configurations. In fact, errors on the wavefront due to cavity defects have been barely discussed in the literature so far when studying the telecentric configuration because their scale has been usually assumed to be large compared to the very small footprint of the incident beam on such etalons. However, etalons are affected not only by large scale defects, but also by high-frequency microroughness or polishing errors of the surfaces, which are usually distributed almost uniformly over the etalon area \citep[e.g.,][]{ref:ibis}. If these defects dominate over other sources of error, the choice of the optimum configuration will depend on their scale relative to the footprint. Telecentric setups minimize wavefront errors if the thickness map vary spatially in a scale larger than (or comparable to) the size of the footprint on the etalon, but, if these variations are of very high frequency, then they could have an important impact on the wavefront. In this case the superiority of telecentric mounts over the collimated configuration in terms of wavefront degradation is not so clear and could be surpassed by the other drawbacks that are present in these setups.

\subsection{Degradation of image quality intrinsic to telecentric mounts}
\label{sec:intrinsic}
Telecentric beams introduce wavefront errors on the mere fact that pupil illumination is no longer homogeneous as seen from the etalon \citep[\emph{pupil apodization,}][]{ref:beckers}. In \citetalias{ref:PaperI} we argued that apodization of the pupil is responsible for a transfer of energy between the central part of the PSF and its wings, thus degrading the image with respect to a perfect unaberrated optical system. This is a wavelength-dependent effect that introduces artificial features in the observed image, as evaluated in \citetalias{ref:PaperIII}, and would occur even in a perfect etalon with no defects.

To estimate the impact of pupil apodization on the wavefront, we can deal with the integration of rays with different incidence angles on the etalon as if it were a  ``defect'' on the illumination compared to a collimated beam. The RMS value of the density distribution of such an error is  given by Eq.~[109] of \citetalias{ref:PaperI}. Then, we can use an approach similar to the one followed in Section \ref{sec:defects} to calculate the perturbation produced by this aperture defect on the transmitted wavefront. The RMS value of the phase error at the maximum of the transmission profile, $\langle \Delta\phi\rangle_{\rm peak}$, is then simply given by

\begin{equation}
\langle \Delta\phi \rangle_{\rm peak}=\frac{1+R}{1-R}\frac{\pi h}{\lambda n\prima (f\#)^28\sqrt{3}}.
\label{eq:wavefront_tele}
\end{equation} 
 In order to estimate the total (monochromatic) degradation of the wavefront produced by a telecentric etalon, this expression should be added quadratically to the perturbation introduced in Eq.~(\ref{eq:propagation}), which accounts for the impact of irregularities. 

 Observe also that a careful choice of the reflectivity is mandatory in telecentric instruments even when defects are ignored, especially for compact instruments with low $f$-numbers. Beams faster than $f/60$ (or $\sim f/140$ in an air-gapped etalon) are almost prohibitive in terms of monochromatic imaging performance even for moderate reflectivities of the order of $\sim 0.9$. In particular, the minimum $f$-number that achieves a wavefront degradation smaller than $\lambda/14$, when no defects are present, is $\sim f/40$ for $R=0.90$ and $\sim f/60$ for $R=0.95$. 

The above results represent a worst-case scenario because image quality has been evaluated monochromatically at the peak of the transmission profile. Once again, we can use Eq.~(\ref{eq:kappa}) to take into account the finite passband of the etalon. This relaxes the diffraction-limiting requirement on the incident beam aperture considerably. In fact, the limiting $f$-number that keeps the RMS error  better than $\lambda/14$ is $\sim f/30$ and $\sim f/40$ for $R=0.90$ and $R=0.95$, respectively. Moreover, much of this degradation can be elliminated by a simple refocus of the etalon \citep{ref:scharmer}. There are compelling reasons to illuminate the etalon with much slower beams, though. As already mentioned, pupil apodization introduces other undesired effects, apart from phase errors, which can be greatly reduced when increasing the $f$-number (\citetalias{ref:PaperIII}). Unfortunately, the larger the $f$-number, the less compact the instrument and the bigger the etalon. Hence, a compromise must be found between artificial signals and the aperture of the incident beam to minimize the effects of pupil apodization while containing the size of the instrument and etalon within realistic and affordable limits. 
To select the optimum aperture of the incident beam on the etalon, we recommend a careful assessment on the impact of the finite aperture of the incident beam taking into account a complete consideration on the polychromatic nature of the observations in the way described in \citetalias{ref:PaperIII}.

\subsection{Discussion on the imaging performance of the two configurations}\label{sec:discussion}
The location of the etalon in a telecentric configuration within the optical path can be chosen carefully to minimize the footprint of the incident beam (and, hence, the impact of high-frequency errors in the wavefront) at the cost of eliminating the possibility of refocusing the etalon in the way described in \cite{ref:scharmer} to reduce phase errors originated by the finite aperture of the incident beam. Yet, wavefront errors produced in this setup are still expected to be smaller than in a collimated configuration.

Unfortunately, if the telecentric configuration is chosen to reduce the impact of defects in the wavefront, then there is also a risk of having different spatial PSFs and transmission profiles across the FoV due to the local variations of the optical thickness over the aperture. This is especially true if two or more etalons are used in tandem to improve the free spectral range and the resolving power of the instrument, each one with a different cavity map. In such a case, the transmission profile is not only shifted, but it becomes asymmetric and its peak is reduced due to the detuning of the individual transmission profiles of each etalon that take place point to point. These effects induce artificial signals in the spectrum of the observed Stokes vector that can be larger than the required polarimetric sensitivity of the instrument locally. To reduce the impact on the Stokes profiles, differential shifts of the spectral profiles must be kept as low as possible by minimizing cavity errors. First order corrections of the measured data are also possible if a careful  reduction technique is followed. An example of a flat-fielding procedure that mitigates successfully the effect of the loss of invariance on the spectral profile can be found in \cite{ref:crispred}. 

Telecentric etalons present other problems that must be considered as well, like artifacts introduced by the strong spectral dependence of their PSF or by deviations from perfect telecentrism (\citetalias{ref:PaperI} and \citetalias{ref:PaperIII}). The latter can arise simply when tilting one of the etalons to move inner etalon ghost images away from the detector. To reduce both effects, the $f$-number of the incident beam should be as large as possible. If two Fabry-P\'erots are employed, it is also highly advisable, first, to combine a low and a high finesse etalons and, second, to apply the minimum necessary tilt only (or mostly) to the etalon with lowest resolution \citep{ref:scharmer}. 

 The collimated configuration is not exempt of problems either, especially when more than one etalon is employed. Differential shifts of the individual transmission profiles  over the FoV can also appear in tandems of collimated etalons when one of them is tilted to avoid ghost images on the detector. The shifts of the individual transmissions across the FoV causes field-dependent asymmetries on the total transmission profile that have the same impact as the ones described above for the telecentric configuration. Moreover, using two collimated etalons in tandem can decrease easily the optical quality of the instrument below requirements due to the amplification of cavity errors, unless they are kept small enough (typically between $0.5$ and $1$ nm RMS over the full clear aperture).

A proper choice on the optimum configuration needs, then, careful considerations on the impact of cavity defects and of tilts of the etalon on the measured signals based on the expected thickness maps of the etalons to be employed. Also important are the implications associated to each configuration on  the required dimensions, quality and costs of the etalons (and of the instrument itself). In particular, the diameter of the etalon in a collimated configuration, $\varnothing_{\rm coll}$, assuming a square $F\times F$ FoV, can be related to the entrance pupil diameter of the telescope, $\varnothing_{\rm pup}$, and to the maximum allowed  spectral shift of the transmission profile across the FoV, $\Delta\lambda$, using the Lagrange invariant and Eq.~[33] of \citetalias{ref:PaperI}, as

\begin{equation}
\varnothing_{\rm coll}=\cfrac{F\varnothing_{\rm pup}}{2n\prima\sqrt{\cfrac{\cal{\Delta\lambda}}{\lambda}}},
\end{equation}
which can be rewritten in terms of the spectral resolving power of the etalon, $\cal{R}$,  and $\epsilon\equiv\Delta\lambda/\delta\lambda$, $\delta\lambda$ being the spectral resolution of the Fabry-P\'erot at the wavelength of interest:
\begin{equation}
\varnothing_{\rm coll}=\frac{F\varnothing_{\rm pup}}{2n\prima}\sqrt{\frac{\cal{R}}{\epsilon}}.
\end{equation}
For the telecentric configuration, making use of the Lagrange invariant  again, the diameter of the etalon, $\varnothing_{\rm tel}$, is given simply as 

\begin{equation}
\varnothing_{\rm tel}=\sqrt{2}F\varnothing_{\rm pup}f\#,
\end{equation}
 where $f\#$ is the $f$-number of the incident beam on the etalon. The ratio of the sizes corresponding to both configurations depend therefore only on the resolving power, the $f$-number, the refraction index, and $\epsilon$, like

\begin{equation}
\frac{\varnothing_{\rm coll}}{\varnothing_{\rm tel}}=\sqrt{\frac{\cal{R}}{8\epsilon}}\frac{1}{n\prima f\#}.
\end{equation}

\begin{figure}
	\centering
	\includegraphics[width=1.05\linewidth]{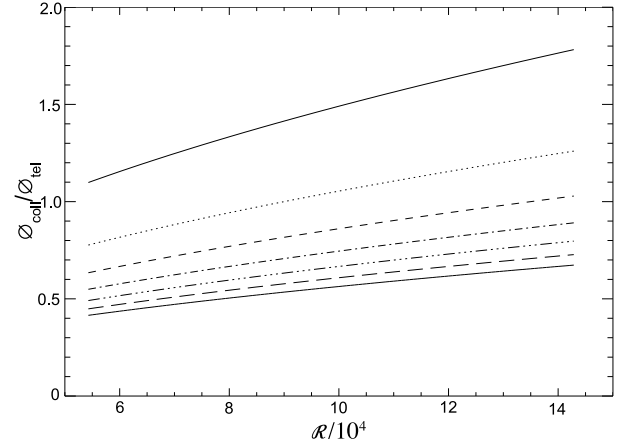}
	\caption{Size ratio of the etalons as a function of the resolving power. Different lines correspond to various values of $\epsilon$. From top to bottom, $\epsilon=0.25,0.5,0.75,1,1.25,1.5,1.75$}
	\label{fig:sizeratio}
\end{figure}

Figure~\ref{fig:sizeratio} shows the ratio $\varnothing_{\rm coll}/\varnothing_{\rm tel}$, parameterized with the value of $\epsilon$,  as a function of $\cal{R}$ for the  case $n\prima f\#=150$. Note that the ratio is below or only slightly above unity for resolving powers up to $\sim150.000$, unless the requirement on the maximum tolerable shift across the FoV is set as tight as $\Delta\lambda=0.25\delta\lambda$. The allowed shift on the collimated configuration differs from one instrument to another, but is usually of the order of $\epsilon=0.75$. This choice of $\epsilon$ guarantees that a maximum of only one wavelength sample is lost at the corner of the FoV when a critical sampling is assumed (i.e., when the spectral sampling is $0.5\delta\lambda$). For this value of $\epsilon$ the ratio is larger than  unity only when resolving powers above $150.000$ are required. 

The next generation of 4 m ground-based telescopes will require etalons with diameters of the order of 150-200 mm or more, no matter the chosen configuration. Requirements in the cavity errors of $\sim 0.5$ nm for such large etalons might by simply unfeasible (or prohibitively expensive), thus eliminating the possibility of using a collimated configuration. Meanwhile, future space-borne telescopes, with much smaller apertures and dimensions, can benefit from the use of collimated mounts for two reasons: first, this setup avoids the problems related with the use of the fast beams that  would be required if a telecentric configuration were employed in such compact instruments and, second, the etalons to be employed in this case are much smaller than those needed in ground-based instruments, making it easier to manufacture them with qualities high enough to ensure diffraction-limited performance.  

\section{Analytical expressions for birefringent etalons}\label{sec:bir}
Electro-optical etalons, like the ones employed in IMaX \citep{ref:imax} and PHI \citep{ref:solanki}, are filled with an anisotropic material that shows birefringent properties. Let us consider a birefringent etalon within a perfect telecentric configuration, where the chief ray is parallel to the optical axis over the whole FoV. The transmission profile is, then, given by Eq.~[50] of \cite{ref:PaperII} , hereinafter Paper II. This expression depends on the Jones matrix terms, which, in turn, depend on the retardances of the ordinary and extraordinary beams, $\delta_{\rm o}$ and $\delta_{\rm e}\equiv \delta_{\rm o}+\varphi$, where $\varphi$ is given by Eq.~[36] of \citetalias{ref:PaperII} . For small incidence angles, we can approximate $\delta_{\rm o}$ and $\delta_{\rm e}$ as

\begin{gather}
\frac{\delta_{\rm o}}{2}=a(1-b_{\rm o}\varrho^2),\\
\frac{\delta_{\rm e}}{2}=a(1-b_{\rm e}\varrho^2),
\end{gather}
where $a$ coincides with Eq.~(\ref{eq:a}), $b_{\rm o}$ is just Eq.~(\ref{eq:b}) with $n\prima=n_{\rm o}$ and $b_{\rm e}$ is given by

\begin{equation}
b_{\rm e}=b_{\rm o}-c,
\end{equation}
where we have defined $c$ as
\begin{equation}
c\equiv\frac{n_3-n_{\rm o}}{n_{\rm o}(n_3+n_{\rm o})^2}\frac{1}{(f\#)^2}.
\end{equation}
Integration of the electric fields for the ordinary and extraordinary rays yields Eq.~(\ref{eq:int_real}) and Eq.~(\ref{eq:int_imag}) with the only difference that $b$ must be substituted with $b_{\rm o}$ or $b_{\rm e}$ correspondingly. Then, an analytical expression for the Mueller matrix of telecentric etalons can be found simply using Eqs.~(45)-(48) of \citetalias{ref:PaperII}. Although the resulting analytical equations are quite laborious and will not be shown here, this method offers an efficient way of calculating the Mueller matrix of etalons in telecentric setups without the need of performing numerical integration, which facilitates calibration and post-processing tasks on space instruments based on anisotropic etalons, whose computational resources are limited.  

\section{Summary and conclusions}
We have solved analytically the equation that governs the transmitted electric field of isotropic telecentric etalons. The found solution is valid for large $f-$numbers ($f\#\gg 1$) typical of solar instruments and is determined only by the reflectivity, absorptivity and two coefficients that are proportional to $n\prima h\lambda^{-1}$ and $(n\prima f\#)^{-2}$, respectively, where $n\prima$ is the refraction index of the etalon, $h$ its thickness and $\lambda$ the wavelength of interest. The fact that $n\prima$ appears only as a proportionality factor of $h$ and of the $f$-number shows that there is a unique equivalence between the solution corresponding to a crystalline etalon and an air-gapped Fabry-P\'erot whose thickness and  $f$-number is  $n\prima$ times larger.  Then, our results obtained for a crystalline etalon with $n\prima=2.3$ and $h=250$ $\mu$m illuminated with telecentric beams ranging from $f/40$ to $f/100$ are completely general also for an air-gapped etalon with $n\prima=1$ and $h=575$ $\mu$m placed in beams with apertures that go from $f/92$ to $f/230$, typical of ground-based solar instruments. This means that crystalline etalons can be placed in much faster telecentric instruments compared to their air-gapped counterparts, with the obvious advantages in the instrument and etalon dimensions.
 
 From the analytical expression of the electric field, we have obtained its derivatives and we have evaluated the sensitivity of the transmission profile and of the phase of the electric field to variations in the etalon parameters. We have shown that the transmission is barely affected by changes in the incident $f$-number, but depends strongly on the thickness, refraction index and wavelength. Similarly to collimated etalons, the transmission profile is mostly shifted by disturbances on the optical cavity, whereas small changes in the $f$-number produce both a shift and a change of width on the profile. 
The phase of the transmitted electric field is also affected by changes in the optical thickness, but in a lesser extent than in collimated instruments. This is due to the lower resolution of the profile in telecentric setups, which translates into a smoother spectral dependence of the phase. The lower the reflectivity and the larger the $f-$number, the more similar the impact to the collimated case.

To account for the limited resolution of the etalon, we have estimated the quasi-monochromatic sensitivity of the phase by integrating its derivative over the transmission profile. We have compared it with the sensitivity at the maximum of the transmission profile through a proportionality factor that depends slightly on the reflectivity and on the aperture of the beam. We have shown numerically that this factor approaches to $\sim 1/2$ for the collimated configuration, as predicted by \cite{ref:scharmer}. For the telecentric case the factor is larger than $1/2$ and increases with decreasing $f$-number, although it approaches to $1/2$ for low to moderate reflectivities.

A simple expression to evaluate the wavefront degradation produced by roughness errors on the etalon surfaces has been presented, too. Such an expression suggests that the choice of the reflectivity plays a very important role in the magnitude of the distortion of the wavefront, setting a limit to the maximum allowed RMS value of the irregularities on the optical thickness over the footprint, as already observed for the collimated configuration by \cite{ref:scharmer}. Telecentric setups are expected to provide a better optical quality than collimated mounts showing the same distribution of defects over the aperture, unless the etalon is mostly affected by thickness errors and/or inhomogeneities of very high frequency that vary in spatial scales smaller than the footprint of the incident beam.
 We have derived also an expression to infer the image degradation that appear in telecentric mounts only because of the finite aperture of the incident beam. Such a degradation has a strong dependence with the reflectivity and with the $f-$number and appears even if no defects are present in the etalon. This effect compensates somehow the lower sensitivity of telecentric etalons to defects, although it can be partially corrected with a refocus of the etalon. 
 
  We have included a discussion on the differences in optical and spectral performance of the telecentric and collimated configurations that accounts for other important effects. We have presented expressions for the size of the etalon in each configuration, too. Telecentric etalons are safer in terms of wavefront distortion, but can introduce artificial signals in the measured magnetic field and line-of-sight plasma velocities, as well as point-to-point variations (and even asymmetries) of the PSF and of the spectral transmission, especially when several etalons are located in tandem. Meanwhile, collimated setups with two or more etalons take the risk of over amplifying the wavefront errors in excess and also introduce field-dependent asymmetries in the transmission if one of the etalons is tilted with respect to the other. The magnitude of these effects must be assessed together with other considerations on the dimensions and cost of the instrument and of the etalon to choose the optimum configuration for each particular instrument.

Finally, we have extended our formulation to the case in which the etalon is anisotropic. In particular, we have introduced a simple modification of the solution valid for the isotropic case that allows for the direct calculation of transmitted electric field corresponding to the ordinary and extraordinary rays. The electric fields can be employed, then, to calculate analytically the Mueller matrices of telecentric etalons in the way described in \citetalias{ref:PaperII} without the need of integrating the equations numerically.

\begin{acknowledgments}
We found the formulas that relate the size of the etalon with the optical parameters of the telescope for the first time in a detailed technical note prepared by Fabio Cavallini in the framework of the SOLARNET project. Our discussion on the size of the etalons is clearly influenced by his note and we would like to publicly thank his contribution. We also owe a debt of gratitude to Göran Scharmer, Francesco Berrilli and Luca Giovannelli for the fruitful debates we have had in the last months on the benefits and drawbacks of each configuration as part of the tasks of the working group on tunable-band imagers for the future European Solar Telescope. Without their contributions, the discussion on the spectral and imaging performance of the collimated and telecentric setups presented in this work would not be as detailed as it is today.
This work has been supported by the
Spanish Science Ministry “Centro de Excelencia Severo
Ochoa” Program under grant SEV-2017-0709 and project
RTI2018-096886-B-C51. D.O.S. also acknowledges financial
support through the Ramón y Cajal fellowship.	
\end{acknowledgments}



\end{document}